\documentclass[11pt,twoside]{book}
\usepackage{vshalo}
\usepackage{longtable}
\usepackage{amsmath,amssymb}
\usepackage{graphicx}
\usepackage{lscape}
\usepackage{epsfig}
\usepackage{index}
\usepackage{natbib}
\usepackage{bigdelim}
\usepackage{multirow}
\makeindex
\newindex{aut}{adx}{and}{\Large Author  Index}
\newcommand{\axindex}[1]{\index[aut]{#1}}

\begin{document}

\pagestyle{myheadings}
\setcounter{equation}{0}\setcounter{figure}{0}\setcounter{footnote}{0}\setcounter{section}{0}\setcounter{table}{0}\setcounter{page}{35}
\markboth{Breger}{The Blazhko Effect in $\delta$~Scuti and other Groups of Pulsating Stars}
\title{The Blazhko Effect in Delta Scuti and other Groups of Pulsating Stars}
\author{Michel Breger}\axindex{Breger, M.}
\affil{}

\begin{abstract} 
Amplitude and period variations have been detected in almost all types of pulsating stars. Many of these modulations are nearly periodic.
In RR Lyrae stars the phenomenon in known as the Blazhko Effect. Because of the observed similarities, we propose to extend the definition to other groups.
We illustrate the Blazhko Effect with examples from Cepheids, RR Lyrae stars, White Dwarfs, sdB stars and $\delta$~Scuti variables.

For $\delta$~Scuti stars the present results indicate the presence of at least two effects: beating of independent modes with close frequencies and stellar cycles. For period and amplitude changes with timescales less than one year, the beating hypothesis explains the observed modulations very well. This has been shown by the correctly correlated relationship between amplitude and phase changes as well as the repetitions of these cycles. However, the observed period variations with longer timescales cannot be due to simple beating between two close frequencies. For the star AI~CVn (= 4~CVn), we can derive accurate annual frequency values for a number of radial and nonradial modes from 1974 to 2009. For prograde and retrograde modes, the frequency variations are of similar size, but with opposite signs. The radial mode shows no (or little) changes. Furthermore, all frequency variations show a reversal around 1990. These results are consistent with long-term cycles affecting individual modes differently with some common systematic behavior. The observed change in the size of rotational splitting is interpreted as a change in the differential rotation during the stellar cycle.
\end{abstract}

\section{Introduction}
The recent ground-based and satellite surveys have shown that amplitude and period variability is common in many different types of pulsating stars. In most cases the timescales and reversals of the effects exclude an evolutionary origin. For the different types of stars, a number of different astrophysical explanations have been offered. These have not been satisfying because of their lack of predictive character as well as their lack of applicability to stars beyond a narrow group or even subgroup. An example might be the resonance hypothesis for RR~Lyrae stars.

In RR Lyrae stars the variations in amplitude, frequency and shape of the light curves is known as the Blazhko Effect, named after Sergei Nikolaevich
 Blazhko, who discovered the variations in RW Dra (Blazhko 1907).
Because the phenomenon is common among so many types of pulsating stars, we would like to expand the definition to include all types of stars, not just RR Lyrae stars. Furthermore, we consider only effects intrinsic to the stars themselves. This means that the external causes of amplitude and frequency variability such as regular light-time effects in binary orbits (e.g., in the $\delta$~Scuti star SZ Lyn, see Derekas et al. 2003) and beating caused by different modes with similar frequencies need to excluded, as far as it is possible to do so from limited observational data. It is hoped that such a wide definition of the Blazhko Effect will motivate researchers to consider the different faces of the effect in different types of pulsators in order to finally reveal the astrophysical processes responsible for these changes.

\section{Cepheids}

While the slow period variations of cepheids are generally compatible with expected evolutionary changes, a number of cepheids have been detected to exhibit much faster and spectacular changes. The most prominent example is Polaris (P = 3.97d), which showed a decrease in amplitude from A$_V$ = 120 to 30 mmag around 1980. The pulsation never stopped completely and the amplitude is increasing again. Bruntt et al. (2008) write, "It now appears that the amplitude change is cyclic rather than monotonic and most likely the result of a pulsation phenomenon", i.e., presumably the Blazhko Effect examined in this paper. Furthermore, around 1965 the smooth (evolutionary?) period change was accompanied by a near-discontinuity in the period changes, a so-called "glitch" (see Turner 2009). We note that the period glitch occurred decades before the amplitude reversal so that the two events may not be directly related.

Shorter timescales of amplitude variability are found for the shorter-period cepheids, e.g., the star V473 Lyr with a pulsation period of 1.49d
(Breger 1981, Burki et al. 1982). For this star the amplitudes vary by about a factor of ten with a Blazhko period near 1200d.

\section{RR Lyrae stars}

RR Lyrae stars are the classical pulsators in which the Blazhko Effect was discovered (Blazhko 1907). The modulation periods are generally short (e.g., 40d) relative to the pulsation periods (e.g., 0.5d), at least when one compares RR~Lyrae stars with other groups of pulsators. It has recently been established that the amplitude variations in RR~Lyrae stars are not exotic or unusual: the excellent work by Jurcsik et al. (2009) at Konkoly Observatory shows that more than 50\% of the stars studied show a detectable Blazhko Effect.

More details are presented in the paper by Katrien Kolenberg in this volume. A very comprehensive analysis of the different physical explanations for the Blazhko Effect can be found in Kov\'acs (2009). The paper confirms that the astrophysical reasons for the observed amplitude and frequency variations are not yet known.

\section{White Dwarfs}

The gravity-mode pulsations of the (evolved) White Dwarfs are also afflicted by severe amplitude variability. An example is the star GD 358 studied by Kepler et al. (2003). Large amplitude changes appear when one compares the Fourier Transforms from 1996 August and September (their Figure 7).
Even though the pulsation amplitudes change on timescales of days and years, the eigenfrequencies remained
essentially the same. This is typical of the Blazhko Effect, rather than mode switching.

\section{sdB stars}

The amplitude and frequency variations of the sdB stars resemble those of the $\delta$~Scuti stars, but are more extreme. A spectacular example is the high-amplitude, hybrid pulsator Balloon 090100001 (Baran et al. 2009). This star is a hybrid pulsator with both gravity (g) and pressure (p) modes. The p modes fall into several distinct frequency regions with similar frequency separations. This is also found in $\delta$~Scuti stars (Breger et al. 2009). The amplitudes and frequencies are strongly variable. Probably the most fascinating result is the increase in the size of the rotational splittings between the 2004 and 2005 observing campaigns, amounting to $\sim$15 per cent. Their results provide strong evidence for differential rotation in the star and a change of this differential rotation within the Blazhko cycle. Below we will provide evidence that this behavior is not unique and that the $\delta$~Scuti stars show similar effects.

\begin{figure}[!h]
\includegraphics*[bb=10 380 650 730,width=165mm,clip]{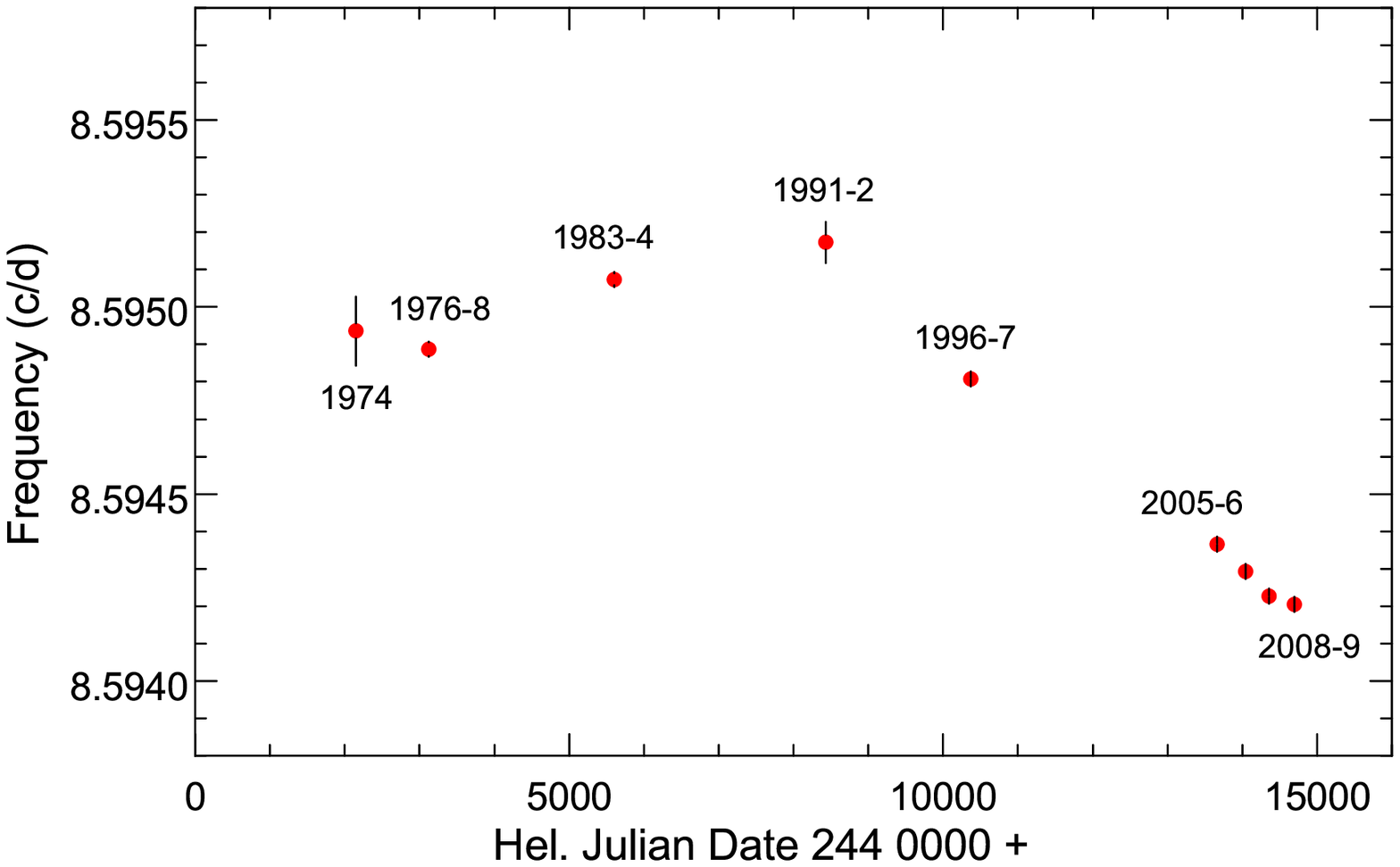}
\caption[]{Frequency variation of the nonradial {\em{prograde}} mode at 8.59 c/d.} 
\label{breger-fig1} 
%
\includegraphics*[bb=10 380 650 730,width=165mm,clip]{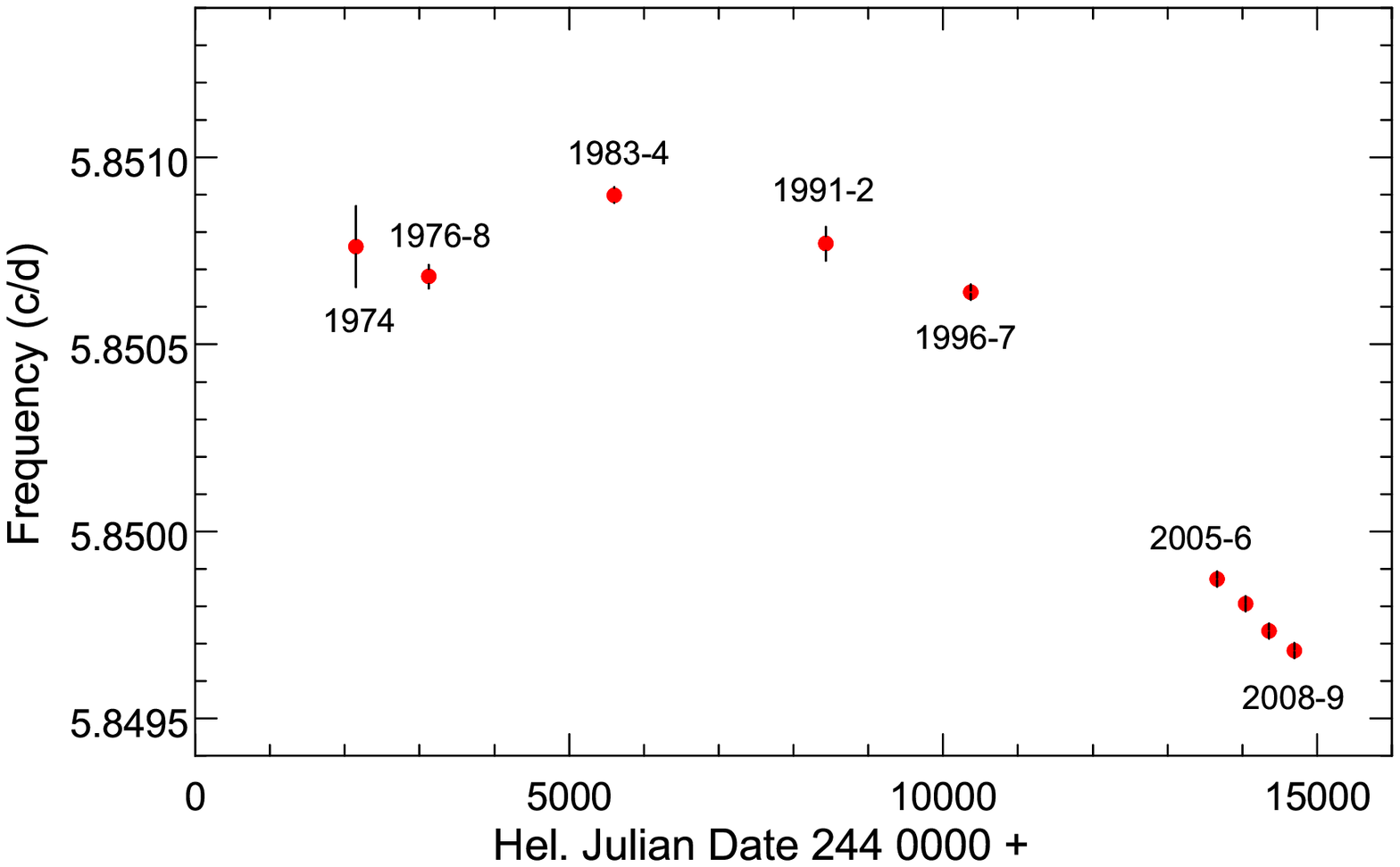}
\caption[]{Frequency variation of another {\em{prograde}} mode (at 5.85 c/d).} 
\label{breger-fig2} 
\end{figure}

\begin{figure}[!h]
\includegraphics*[bb=10 380 650 730,width=165mm,clip]{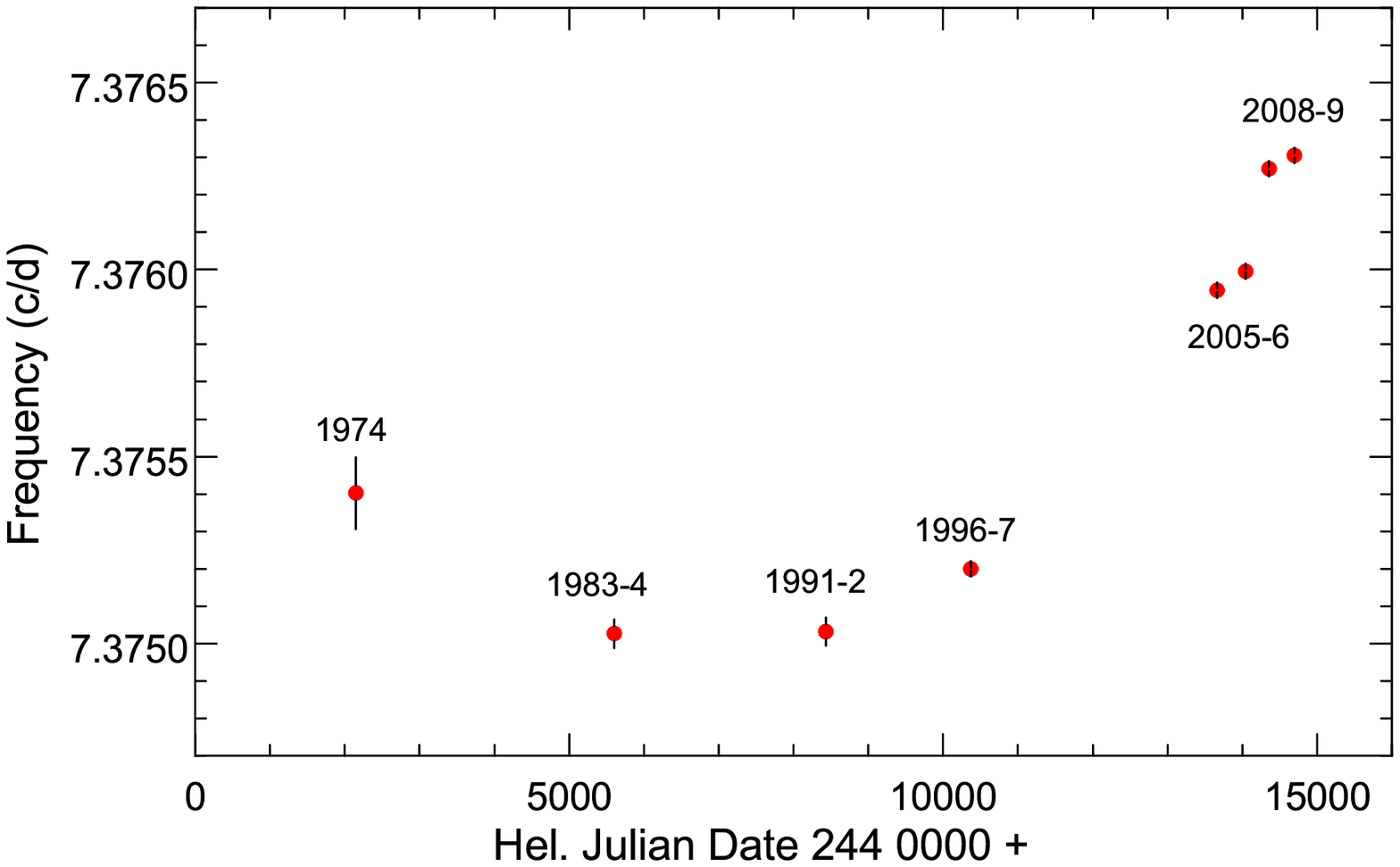}
\caption[]{Frequency variation of the nonradial {\em{retrograde}} mode at 7.38 c/d. } 
\label{breger-fig3} 
\includegraphics*[bb=10 380 650 730,width=165mm,clip]{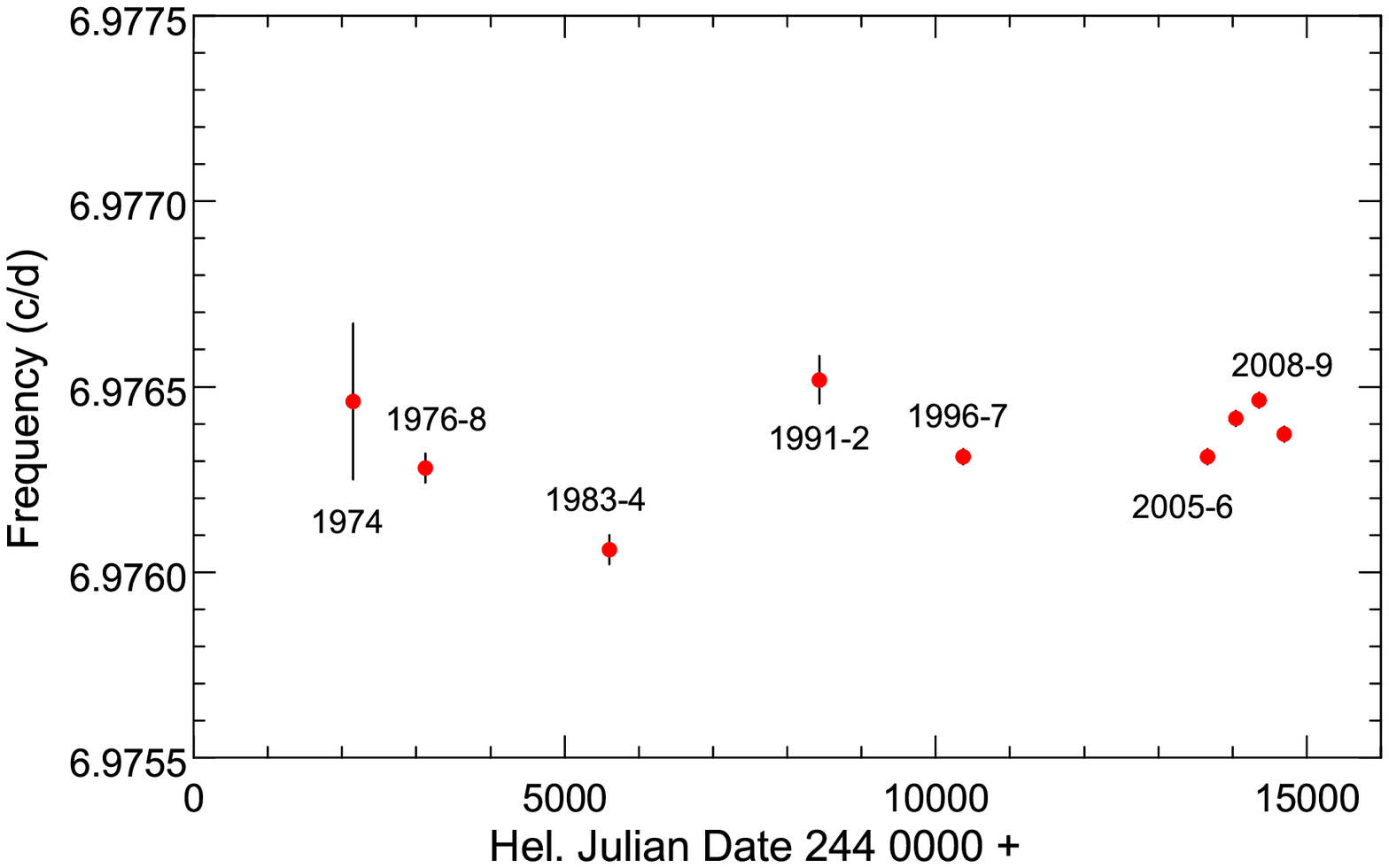}
\caption[]{Frequency variation of the {\em{axisymmetric}} mode at 6.98 c/d. } 
\label{breger-fig4} 
\end{figure}

\section{$\delta$~Scuti stars}

The $\delta$~Scuti stars are A and F stars situated on and above the main sequence. Amplitude variability and period variability is common among  these pulsating stars. Except for the slowly rotating radial pulsators with high amplitudes (the so-called HADS), most variables are nonradial pulsators with small amplitudes in the millimag range. A few variables are extremely well-studied so that considerable information on the Blazhko modulations is available. An example among the HADS is V1162 Ori (Arentoft \& Sterken 2002). 

\subsection{Timescales under one year: beating between close frequencies}

In $\delta$~Scuti stars, the observed amplitude and frequency modulations can be divided into two groups: those with timescales shorter than about 250d and those with longer timescales.

From detailed examinations of the modulations in the well-studied stars FG~Vir and BI CMi, we determined that the observed amplitude and frequency variations with timescales less than about 250d are caused by the beating between two close frequencies. The observed amplitude and phase variations of a number of frequencies with amplitude and phase variability are exactly as predicted from a two-frequency solution; the agreement even repeats from beat cycle to beat cycle (Breger \& Bischof 2002, Breger \& Pamyatnykh 2006).

Until recently, the large number of detected close-frequency pairs seemed difficult to explain, unless resonance effects are involved. The individual modes making up these pairs should be radial or nonradial  $\ell$ = 1 and 2 modes, because  the observable photometric amplitudes of high-$\ell$ modes are too small to cause substantial beating effects in the light curves. However, we now know that in $\delta$~Scuti stars the frequencies of  $\ell$ = 1 and 2 modes are not randomly distributed.
In $\delta$~Scuti stars radial and nonradial modes tend to cluster around the radial modes due to trapped modes in the envelope (Breger, Lenz \& Pamyatnykh 2009). This, in turn, can at least partially explain the large number of close frequencies in $\delta$~Scuti stars.

With insufficient observational data, the beating between two modes of similar frequencies appears as true amplitude and period variability, but is not a true astrophysical modulation of a pulsation mode. Consequently, for studies of the Blazhko Effect in $\delta$~Scuti stars we need to examine the modulations with times scales of longer than about 250d.

\subsection{Longer modulation timescales: The $\delta$~Scuti star AI~CVn}

\begin{figure}[!h]
\includegraphics*[bb=10 60 650 780,width=165mm,clip]{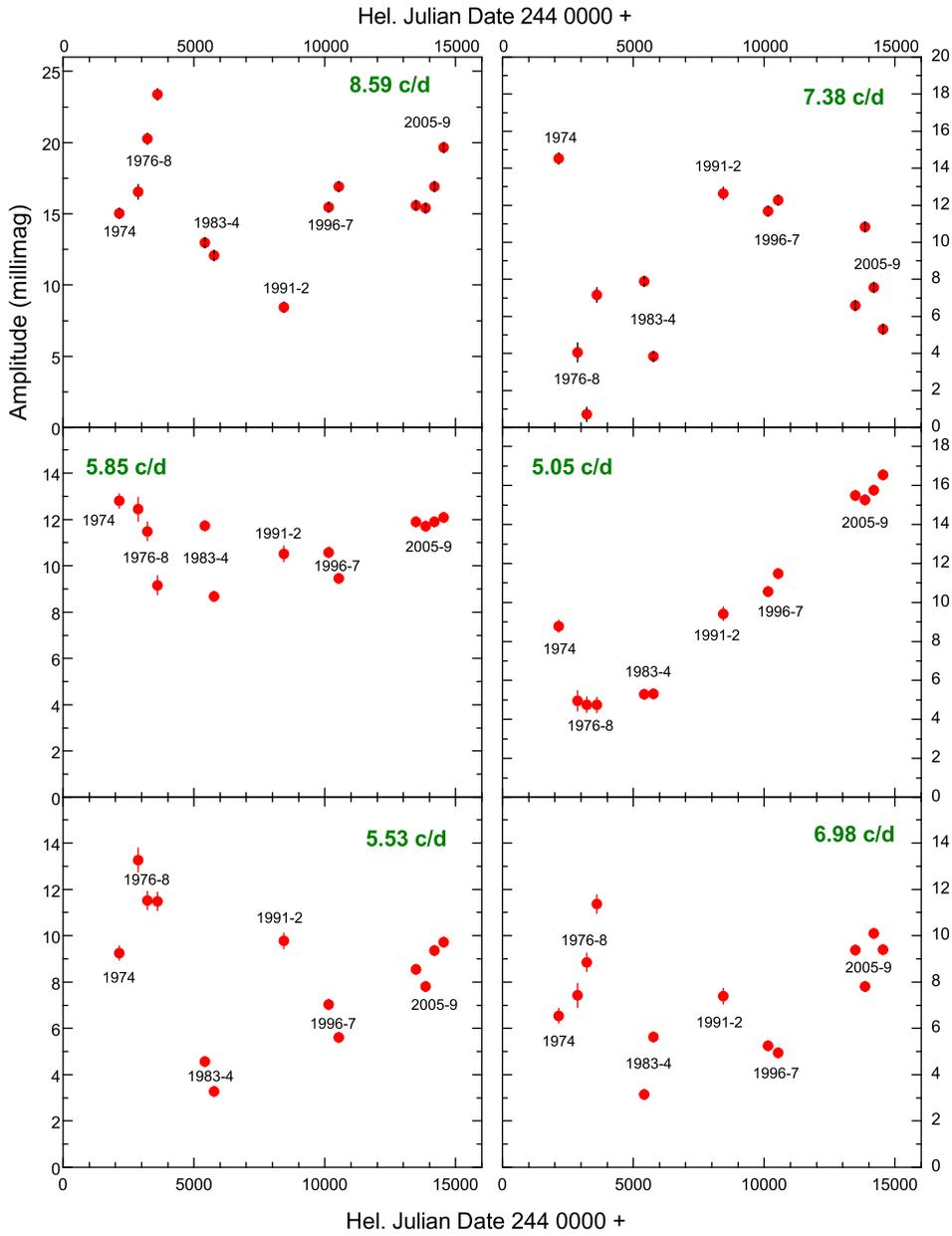}
\caption[]{Amplitude variations of six modes. The figure shows that the amplitude changes are not related to the frequency variations shown in the previous figures.} 
\label{breger-fig5} 
\end{figure}

The star AI~CVn (= 4 CVn) is an evolved $\delta$~Scuti star with known amplitude variability of its many nonradial and radial modes (Breger 2000). It is well-suited for a more detailed study because prograde, axisymmetric as well as retrograde modes have been identified. Consequently, it is possible to study the Blazhko Effect in a variety of different pulsation modes present in the same star.

AI~CVn has been observed photometrically from 1966 up to 2009 (and more measurements are planned for 2010). In order to understand the amplitude and period variations, a telescope was dedicated to study AI~CVn. From 2005 to 2009, the Vienna Automatic Photoelectric Telescope (APT) in Mount Washington, Arizona, USA was utilized for five months per year to measure AI~CVn relative to two comparison stars in two colors. The multifrequency analysis with PERIOD04 (Lenz \& Breger  2005) led to the determination of 72 significant frequencies, of which 60 had small amplitudes of 0.001 mag or less. The two-color photometry together with simultaneous spectroscopic observations (Castanheira et al. 2008) made it possible to carry out a preliminary pulsation mode identification of the frequencies with the largest amplitudes.

The long coverage within each year of the 2005--2009 data showed that the amplitude and frequency variations occurred on long timescales of years. The only exception is the well-resolved close-frequency pair at 6.1170 and 6.1077 c/d, which leads to beating in short data sets (see previous subsection). 

The standard method to study frequency changes involves the calculation of (O-C) phase shifts, i.e., the difference between the observed and predicted times of maximum
when adopting a suitable frequency value. However, for AI~CVn it turned out that this method is not suitable due to longer time gaps between the observations and the sign reversal of the frequency changes near the year 1990 (see below). Fortunately, the data sets are extensive enough to derive annual (or two-year) frequency values independent of the measurements from other years. The following additional data sets could be used:  the extensive unpublished Fitch measurements from 1974 to 1978, and the multisite $\delta$~Scuti Network campaigns from 1983/4 and 1996/7. Furthermore, observations from 1991/1992 have also recently become available (Breger et al. 2008).

Figures 1 through 4 show the frequency variations of four of the 'large-amplitude' frequencies with different values of the azimuthal quantum number $m$. From these results we can conclude the following:

(i) the nonradial prograde modes show an increase in frequency until around the year 1990, followed by a decrease,

(ii) the nonradial retrograde mode shows the exact opposite behavior, while

(iii) the radial (axisymmetric) mode shows only small or zero frequency changes.

Figure 5 shows the changes of amplitude for six modes. The amplitudes are variable with timescales of years or decades (e.g., see the steady change of the 5.05 c/d mode). However, the amplitude changes are not synchronized with the either the frequency changes of the same mode or the amplitude changes of other modes.

We conclude that the frequency changes of the different modes are strongly related. 
The fact that the changes of the prograde and retrograde modes are opposite to each other and that the axisymmetric mode changes little indicates that the {\bf rotational splitting is changing}. Although the observed rotational splitting is mainly a geometric effect caused by stellar rotation (the $m\Omega$ term), this does not necessarily mean that the rotation of the star is changing. It does mean that there exists differential rotation. Nonradial modes of different azimuthal ($m$) values occur at different latitudes, while different $\ell$ values probe different depths inside the star. We interpret these results as a change in differential rotation during the stellar cycle.

\acknowledgements

Part of this work has been supported by the Austrian Fonds zur F\"orderung der wissenschaftlichen Forschung.

\vspace{5mm}
\noindent
\textbf{References:}\\ \\
Arentoft, T., \& Sterken, C., 2002, ASP Conf. Ser., 256, 79\\
Baran, A., Oneira, R., Pigulski, A., 2009, MNRAS, 392, 1092\\
Blazhko, S. N., 1907, Astron. Nachr., 173, 325\\
Breger, M., 1981, \apj , 249, 666\\
Breger, M., 2000, MNRAS, 313, 129\\
Breger, M., \& Bischof, K. M., 2002, A\&A, 385, 537\\
Breger, M., \& Pamyatnykh, A. A., 2006, MNRAS, 368, 571\\
Breger, M., Davis, K. A., \& Dukes, R. J., 2008, CoAst, 153, 63\\
Breger, M., Lenz, P., \& Pamyatnykh, A. A., 2009, \mnras, 396, 291\\
Bruntt, H., Evans, N. R., Stello, D., \& et al. 2008, \apj, 683, 433\\
Burki, G., Mayor, M., \& Benz, W., 1982, A\&A, 109, 258\\
Castanheira, B., Breger, M., Beck, P. \& et al., 2008, CoAst 157, 124\\
Derekas, A., Kiss, L. L., Sz\'ekely, P., \& et al., 2003, A\&A, 402, 733\\
Jurcsik, J., S\'odor, \'A., Szeidl, B., et al., 2009, MNRAS, 400, 1006 \\
Kepler, S. O.,  Nather, R. E., Winget, D. E., \& et al. 2003, A\&A, 401, 639\\
Kov\'acs, G., 2009, AIPC, 1170, 261\\
Lenz, P., \& Breger, M., CoAst, 146, 53\\
Turner, D. G., 2009, AIPC, 1170, 59\\

\end{document}